\documentclass[english,aps,prl,twocolumn,groupedaddress,showpacs]{revtex4}
\usepackage[T1]{fontenc}
\usepackage[latin1]{inputenc}
\usepackage{graphicx}
\usepackage{amssymb}
\usepackage{bm}



\usepackage{babel}
\makeatother

\begin{document}

\title{Decoherence by Quantum Telegraph Noise}

\author{Benjamin Abel}

\affiliation{Department Physik and Center for Nanoscience and Arnold Sommerfeld
Center for Theoretical Physics, Ludwig-Maximilians Universit\"at M\"unchen,
Theresienstr. 37, 80333 M\"unchen, Germany}

\author{Florian Marquardt}

\affiliation{Department Physik and Center for Nanoscience and Arnold Sommerfeld
Center for Theoretical Physics, Ludwig-Maximilians Universit\"at M\"unchen,
Theresienstr. 37, 80333 M\"unchen, Germany}

\date{\today}

\begin{abstract}
We investigate the time-evolution of a charge qubit subject to quantum
telegraph noise produced by a single electronic defect level. We obtain
results for the time-evolution of the coherence that are strikingly
different from the usual case of a harmonic oscillator bath (Gaussian
noise). When the coupling strength crosses a certain temperature-dependent threshold, we
observe coherence oscillations in the strong-coupling regime. Moreover,
we present the time-evolution of the echo signal in a spin-echo experiment.
Our analysis relies on a numerical evaluation of the exact solution
for the density matrix of the qubit. 
\end{abstract}


\maketitle

The unavoidable coupling of any quantum system to a noisy environment
leads to decoherence. Understanding decoherence is interesting for
fundamental reasons (the quantum-classical crossover, the measurement
problem etc.), and is essential for achieving the long dephasing times
neccessary for building a quantum computer and other applications.
The paradigmatic models in this field (Caldeira-Leggett and spin-boson
model \citep{1981_01_CaldeiraLeggett_TunnelingWithDissipation,
1983_CaldeiraLeggett_QuantumBrownianMotion,1987_Leggett_ReviewSpinBoson,
2000_Weiss_QuantumDissipativeSystems,2002_Breuer_Book})
usually consider a bath of harmonic oscillators. In that case, the
bath variable coupling to the quantum system displays Gaussian-distributed
fluctuations. This feature affords considerable technical simplifications,
while these models are faithful descriptions of real environments
like the vacuum electromagnetic field or the harmonic crystal lattice.
In other cases (like electronic Nyquist noise in a bulk metal), these
models represent very good approximations. This is a consequence of
the central limit theorem, applied to the sum of contributions from
many independent non-Gaussian noise sources. The approximation finally
breaks down when one couples strongly to a few noise sources. This
situation is becoming more prevalent nowadays, as one studies the
coherent dynamics of nanostructures. The coherence times of solid
state qubits are often determined by a few fluctuators 
\citep{2004_08_Martinis_JunctionResonators,
2004_12_Astafiev_StrangeNoiseSpectrum, 2007_06_Clarke_FluxNoiseModel}.

This challenge has given rise to a number of theoretical studies of
qubits subject to fluctuators producing telegraph noise \citep{Faoro:2005fk,2005_08_Lerner_QuantumTelegraphNoiseLongTime,shnirman2005lah,2006_03_GalperinAltshuler_NonGaussianQubitDecoherence,desousa2005oas,
bergli2006dqn,2006_01_SchrieflShnirman_DecoherenceTLS,bergli2007esd,galperin2007ngd,
gurvitz2008qma,emary_quantumdynamics_2008} (and other non-Gaussian baths  \citep{shao1998ddt,shao1998sqc,2000_ProkofevStamp_ReviewSpinBath,2002_Gassmann_NonlinearBath,paladino2004mdd,paladino2008cci}).
The most straightforward but realistic fully quantum-mechanical model
consists of a single level tunnel-coupled to an electron reservoir \citep{2002_06_PaladinoFazio_TelegraphNoisePRL}. 
Grishin, Yurkevich and Lerner recently studied the long-time limit
of this model and derived the dephasing rate for a qubit coupled to
such a fluctuator \citep{2005_08_Lerner_QuantumTelegraphNoiseLongTime}.
They found a striking non-analytic dependence of the dephasing rate
on the coupling strength and temperature. In the present paper, we
take up the same model, which may reasonably be termed ``quantum
telegraph noise'', now asking for the full time-dependence. We find
that in the strong-coupling regime (beyond a certain threshold) the
 monotonous decay of the qubit's coherence turns into temporal oscillations,
with complete loss of coherence interspersed between coherence revivals.
We are able to fully include quantum fluctuations, by a numerical
evaluation of the exact solution for the quantum model, and we discuss
the behaviour at low temperatures. We conclude by showing how to extend
these calculations to spin-echo experiments, relevant for coherence
control. 
\begin{figure}
\centering \includegraphics[width=0.99\columnwidth]{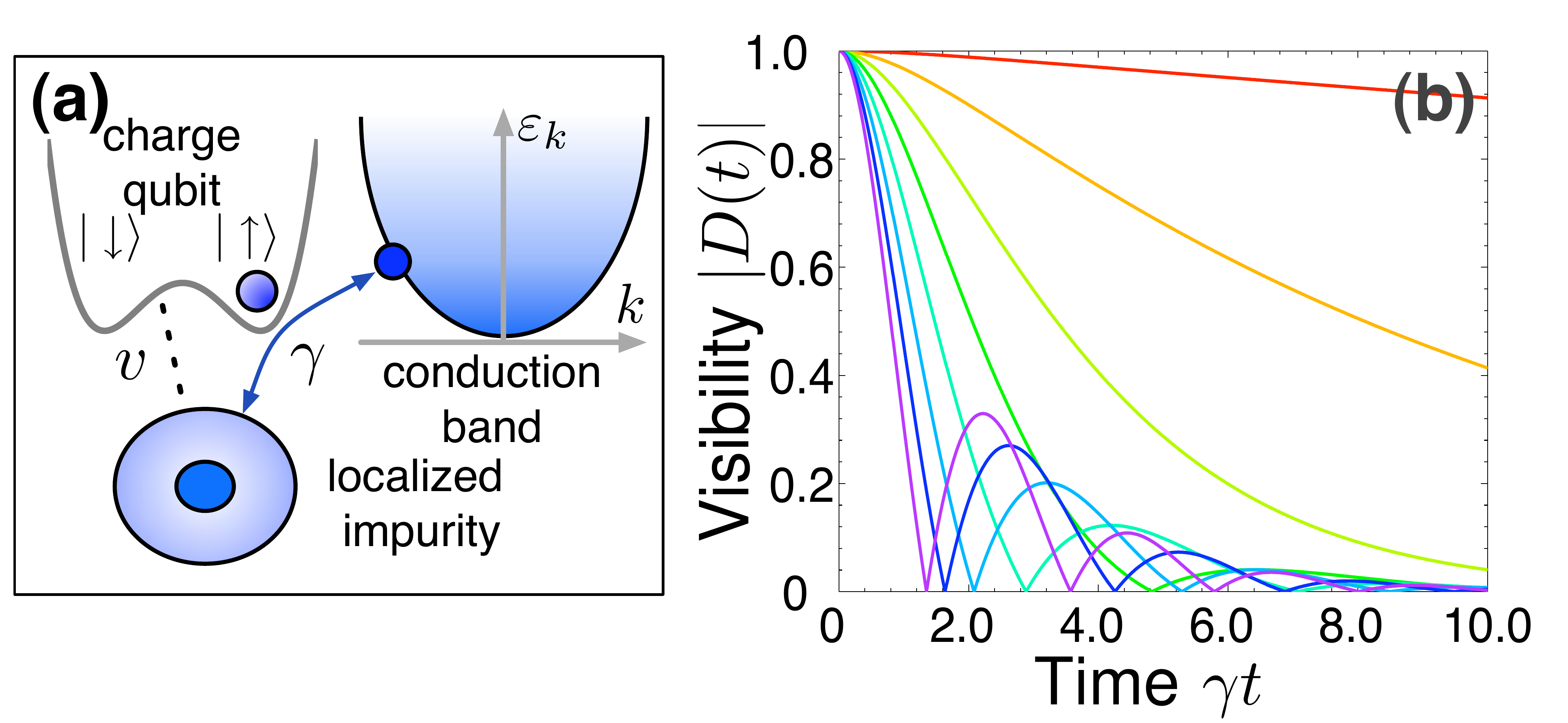}
\caption{(a) Schematic picture of the bistable fluctuator: A localized level
tunnel-coupled to an electron reservoir. (b) Time-evolution of the
visibility $|D(t))|$ for \emph{classial} telegraph noise (top to bottom: $v/\gamma=0.2,\mbox{ }0.6,\mbox{ }1.0,\mbox{ }1.4,\mbox{ }1.8,\mbox{ }2.2,
\mbox{ }2.6,\mbox{ }3.0$).}
\label{fig:classicalcoherence} 
\end{figure}

\textit{Model}.\textendash{} We study a single, spin-polarized impurity
level (Fig.~\ref{fig:classicalcoherence}(a)), tunnel-coupled to a (non-interacting)
electron reservoir: 
\begin{equation}
{\hat{\mathcal{H}}_{B}}=\varepsilon_{0}\hat{d}^{\dag}\hat{d}+\sum_{\bm{k}}
\left(t_{\bm{k}}\hat{c}_{\bm{k}}^{\dag}\hat{d}+\mbox{h.c.}\right)+
\sum_{\bm{k}}\varepsilon_{\bm{k}}\hat{c}_{\bm{k}}^{\dag}\hat{c}_{\bm{k}}.
\label{eq:fluctuator}
\end{equation}
Here $\hat{d}^{\dagger}$ creates an electron on the impurity level
of energy $\varepsilon_{0}$, and $t_{\bm{k}}$ is the tunneling amplitude
to the reservoir level $k$, of energy $\varepsilon_{k}$ (we fix
the reservoir's chemical potential as $\mu=0$). Below, we always
refer to the tunneling rate $\gamma=2\pi\sum_{k}\left|t_{k}\right|^{2}\delta(\epsilon_{k}-\epsilon_{0})$.
The fluctuating impurity charge $\hat{\mathcal{Q}}=\hat{d}^{\dagger}\hat{d}$
couples to a qubit, and the full Hamiltonian is given by ($\hbar=1$,
$k_{B}=1$) 
\begin{equation}
\hat{\mathcal{H}}=\frac{\Delta}{2}\hat{\sigma}_{z}+
\frac{v}{2}\hat{\mathcal{Q}}\hat{\sigma}_{z}+{\hat{\mathcal{H}}_{B}},
\label{eq:Hamiltonian}
\end{equation}
where $\hat{\sigma}_{x,y,z}$ are the qubit Pauli operators, $\Delta$
is the qubit level spacing, and $v$ is the qubit-fluctuator coupling
strength. The coupling considered here leads only to pure dephasing
and not to energy relaxation in the qubit. This is a popular and realistic
model when discussing the decay of quantum information during storage. 

We are interested in the full time dynamics of the reduced density-matrix
$\hat{\rho}(t)$ of the qubit, after preparing it in a superposition
state and switching on the interaction with the fluctuator. Since
the interaction, $\hat{\mathcal{H}}_{{\rm {int}}}=\frac{v}{2}\hat{\mathcal{Q}}\hat{\sigma}_{z}$,
commutes with the qubit Hamiltonian, only the off-diagonal elements
$\rho_{ij}$ are affected ($i,j\in\{\uparrow,\downarrow\}$), acquiring
an additional coherence factor $D(t)$: 
\begin{equation}
\rho_{\uparrow\downarrow}(t)=\rho_{\uparrow\downarrow}(0)e^{-i\Delta t}D(t).
\label{eq:densitymatrix}
\end{equation}
\textit{Classical telegraph noise}. \textendash{} We first review
the classical limit for the bath, where the charge $Q(t)$ is a stochastic
process of the ``telegraph noise'' type \citep{goychuk1995ked},
which flips randomly between $0$ and $1$ (occuring with equal prababilities) at a rate $\gamma$. This
corresponds precisely to the high-temperature limit of the quantum
model discussed here (see below). For a given realization of $Q(t)$,
the Schr\"odinger-equation yields a superposition of the qubit's eigenstates
with a random contribution to the relative phase, $\varphi(t)=-v\int_{0}^{t}dt'Q(t')$.
The noise average yields the coherence, $D(t)=\left\langle e^{i\varphi(t)}\right\rangle $.
If the phase were Gaussian distributed, then the coherence would be
determined by the variance of $\varphi$: $\left\langle e^{i\varphi(t)}\right\rangle 
= e^{i\left\langle \varphi(t)\right\rangle -\frac{1}{2}\left\langle \varphi^{2}(t)\right\rangle }$
. This is not true for classical telegraph noise, where the exact
result is found to be $D(t) = e^{-\frac{i}{2}(v-i\gamma)t} \left[\cosh(\delta t) + (\gamma/2\delta)\sinh(\delta t)\right]$,
where $\delta=\frac{1}{2}\sqrt{\gamma^{2}-v^{2}}$, and $\gamma^{-1}$
is the charge correlation time: $\langle\delta Q(t)\delta Q(0)\rangle=\frac{1}{4}e^{-\gamma|t|}$,
with $\delta Q(t)=Q(t)-\langle Q(t)\rangle$. The {}``interference
contrast'' of any observable sensitive to the relative phase between
the qubit's levels is reduced by the factor $|D(t)|$, which we will
term the visibility. Fig.~\ref{fig:classicalcoherence}(b) shows $|D(t)|$
for different couplings $v$. Coherence oscillations appear when $v>\gamma$,
as $\delta$ becomes imaginary. These are qualitatively different
from anything observed for Gaussian noise, where $D(t)$
cannot cross zero. The long-time decay rate of $|D(t)|$ is equal
to $\frac{1}{2}(\gamma-\sqrt{\gamma^{2}-v^{2}})$ if $v\leqslant\gamma$
and $\gamma/2$ if $v>\gamma$.

\textit{General exact solution}. \textendash{} In the full quantum
model [Eqs.~(\ref{eq:Hamiltonian}) and (\ref{eq:fluctuator})]
the coherence can generally \citep{1990_04_SAI} be written as an
overlap, $D(t)=\langle\chi_{B}^{\downarrow}(t)|\chi_{B}^{\uparrow}(t)\rangle$,
of the two bath-states $|\chi_{B}^{\uparrow}(t)\rangle$ and $|\chi_{B}^{\downarrow}(t)\rangle$
produced under the action of the qubit being in state $\left|\uparrow\right\rangle $
or $\left|\downarrow\right\rangle $. Then the coherence is 
\begin{equation}
D(t)=\left\langle e^{i(\hat{\mathcal{H}}_{B}-
\frac{v}{2}\hat{\mathcal{Q}})t}e^{-i(\hat{\mathcal{H}}_{B}+\frac{v}{2}\hat{\mathcal{Q}})t}\right\rangle ,\label{eq:coherence}
\end{equation}
where we average over the thermal state of the electron bath. A variety
of methods have been applied to calculate averages of the form Eq.~(\ref{eq:coherence}),
e.g. linked-cluster expansions or nonequilibrium Keldysh path-integral
techniques \citep{2004_04_MakhlinShnirman_DephasingOptimalPoints,
2005_08_Lerner_QuantumTelegraphNoiseLongTime}. 
Here we implement a variant of a formula known from full-counting
statistics \citep{1996_06_LevitovLeeLesovik_FullCountingStatistics,2002_Klich_Formula,snyman2007pcq,hassler2008wpf},
which can be evaluated numerically efficiently. Given arbitrary single-particle
operators $\hat{A},\hat{B}$ and $\hat{C}$, and their second quantized
counterparts $\hat{\mathcal{A}}=\sum_{k,k'}\hat{c}_{k'}^{\dagger}A_{k'k}\hat{c}_{k}$
etc., the trace ${\rm tr}[e^{\hat{\mathcal{A}}}e^{\hat{\mathcal{B}}}e^{\hat{\mathcal{C}}}]$
over the many-body Hilbert space is equal to ${\rm det}[1+e^{\hat{A}}e^{\hat{B}}e^{\hat{C}}]$.
Applying this to Eq.~(\ref{eq:coherence}), we obtain 
\begin{equation}
D(t)=\det\left(1-\hat{n}+e^{i(\hat{H}_{B}-\frac{v}{2}\hat{Q})t} 
e^{-i(\hat{H}_{B}+\frac{v}{2}\hat{Q})t}\hat{n}\right).
\label{eq:determinant}
\end{equation}
Here $\hat{H}_{B}$ and $\hat{Q}$ are the single-particle operators
corresponding to $\hat{\mathcal{H}}_{B}$ and $\hat{\mathcal{Q}}$,
and $\hat{n}=f(\hat{H}_{B})$ is the single-particle equilibrium density
matrix, where $f(\varepsilon)=(\exp(\beta\varepsilon)+1)^{-1}$ is
the Fermi-Dirac distribution. This formula takes into account exactly
the effects of quantum fluctuations (on top of thermal ones), and
the non-Markovian features in the fluctuator dynamics that develop
for decreasing temperatures.

\textit{Numerical evaluation}.\textendash{} Our results for the time-evolution
of the visibility have been obtained by direct numerically exact evaluation
of Eq.~(\ref{eq:determinant}). To this end, we employ a discretization
with $N$ equally spaced energy levels $\varepsilon\in[-W,W]$ in
a band $W\gg\gamma$. These represent the single-particle energy eigenlevels
of $\hat{H}_{B}$, for which the matrix-elements of $\hat{Q}$ are
equal to 
\begin{equation}
\hat{Q}_{\alpha\beta}=\frac{1}{\pi\nu}\sqrt{\mbox{Im}G^{R}(\omega = 
\varepsilon_{\alpha})\mbox{Im}G^{R}(\omega=\varepsilon_{\beta})}.
\end{equation}
Here $G^{R}(\omega)=(\omega-\varepsilon_{0}+i\gamma/2)^{-1}$ is the
impurity level's retarded Green's function and $\nu=N/(2W)$ is the
level density. The coherence is obtained by calculating the determinant
of the resulting $N\times N$-matrix, Eq.~(\ref{eq:determinant}).
Good convergence is obtained already for $N$ on the order of $400$
and $W=20$.

\begin{figure}[t]
 \centering \includegraphics[width=0.99\columnwidth]{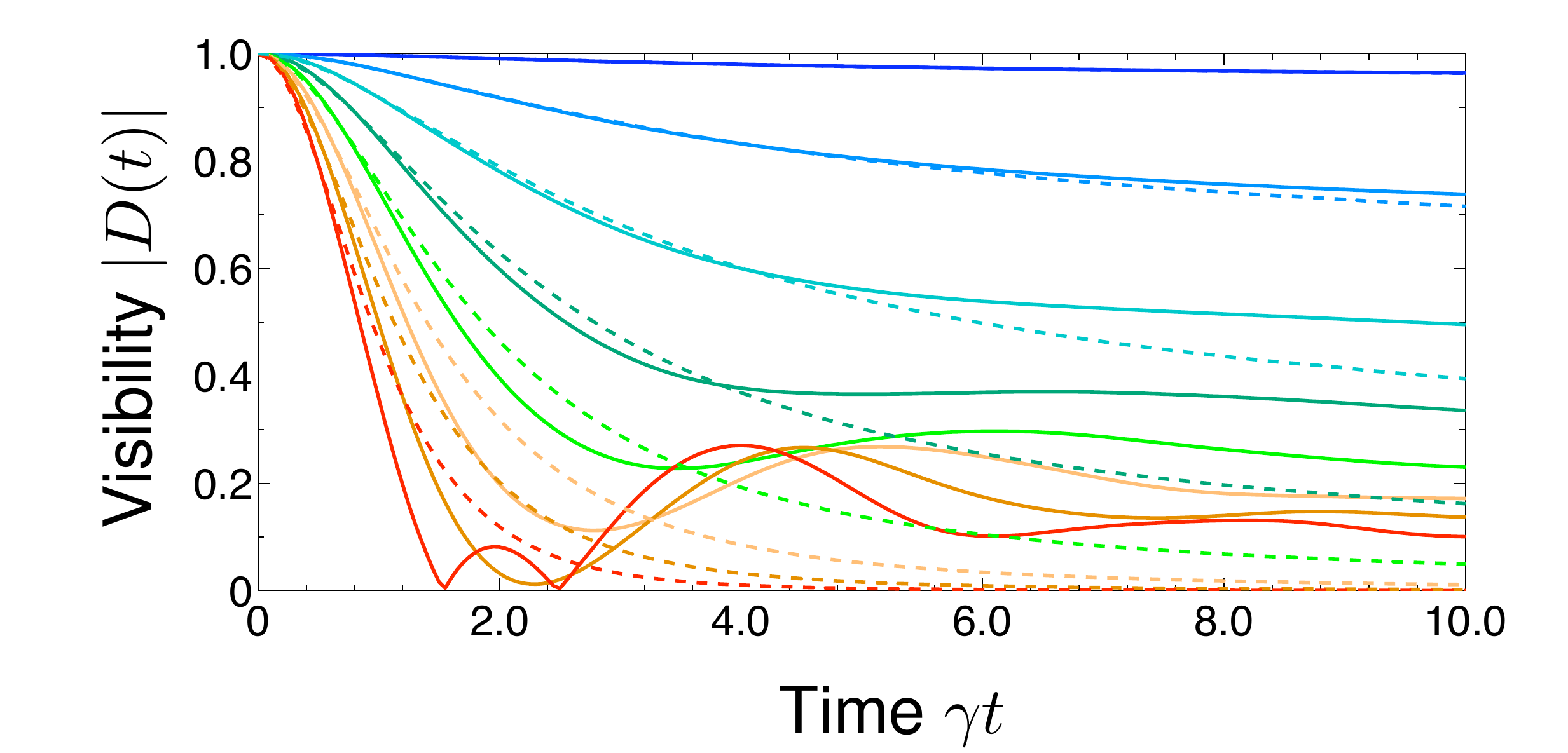} 
\caption{Time-evolution of the visibility $|D(t)|$ for different couplings
$v$, for quantum telegraph noise acting on a qubit at low temperatures
($T/\gamma=0.01$). The dashed lines show the Gaussian approximation.
From top to bottom: $v/\gamma=0.2,\mbox{ }0.6,\mbox{ }1.0,\mbox{ }1.4,\mbox{ }1.8,\mbox{ }2.2,\mbox{ }2.6,\mbox{ }3.0$
(with $\varepsilon_{0}=0$). }
\label{fig:quantumcoherence} 
\end{figure}

\begin{figure}[t]
 \centering 
 \includegraphics[width=0.99\columnwidth]{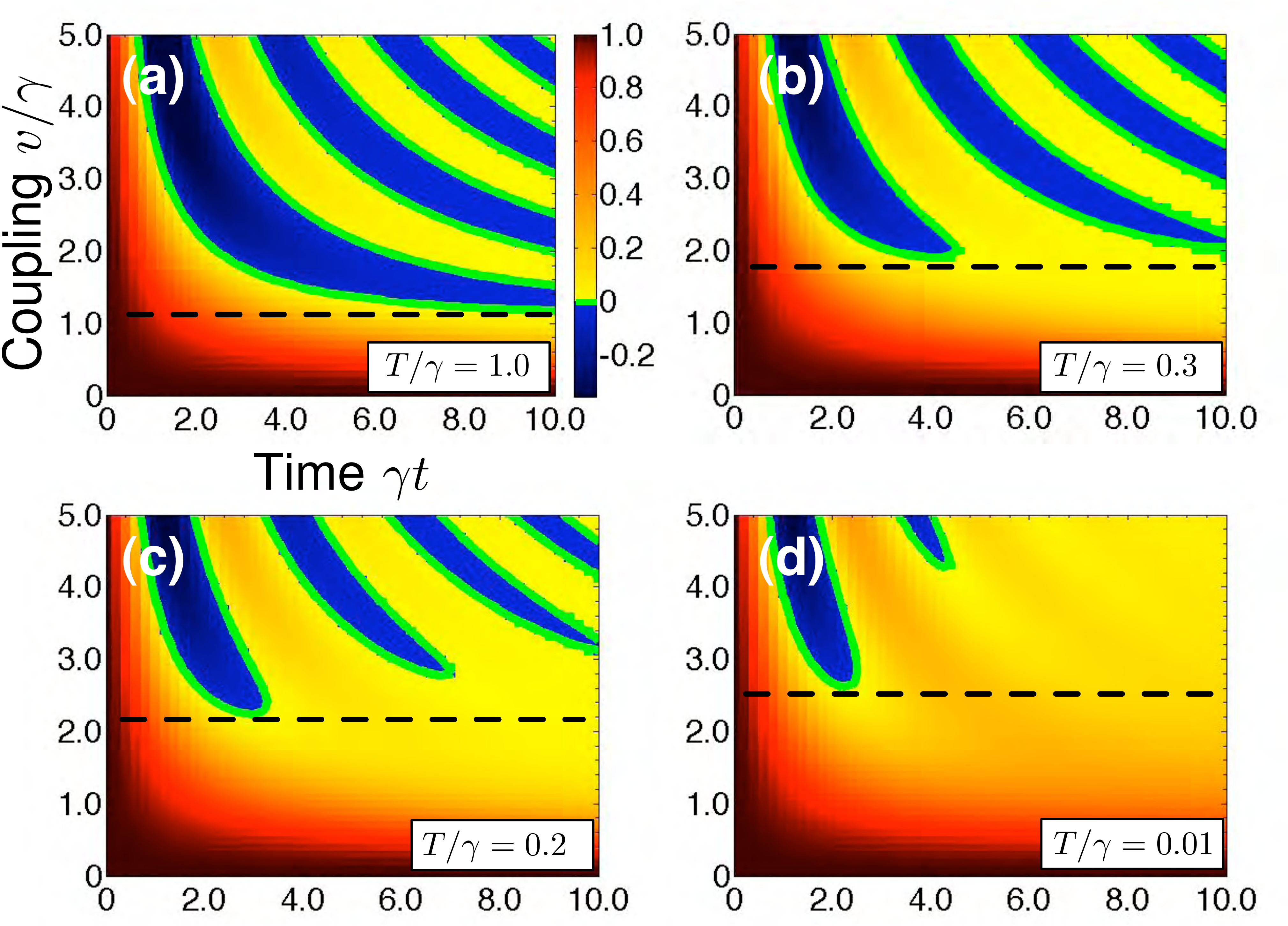} 
\caption{(color online) Density plot of $e^{+i\frac{v}{2}t}D(t)$, which is
real-valued, as a function of time (horizontal) and coupling (vertical).
(a)-(d) $T/\gamma=1.0$, $0.3$, $0.2$, $0.01$ (with $\varepsilon_{0}=0$).
The bold green lines indicate the contours of vanishing coherence,
$D(t)=0$. The dashed line indicates the critical coupling strength
$v_{c}^{q}/\gamma=1.1,\mbox{ }1.9,\mbox{ }2.2,\mbox{ }2.7$ (from
(a)-(d)).}
\label{fig:densityplot} 
\end{figure}

\textit{Results for the visibility}.\textendash{} In Fig.~\ref{fig:quantumcoherence}
we show the visibility for different couplings $v$. For small coupling,
$v/\gamma\ll1$, the Gaussian approximation works well. It can be
obtained from Eq.~(\ref{eq:determinant}) by writing 
${\rm det}(\hat{A})={\rm exp}({\rm tr}({\rm ln}(\hat{A})))$,
and keeping only the terms up to order $v^{2}$ in the exponent (see
also \cite{neder2007cod}). Equivalently, one may use $D(t)$
that would be obtained for a harmonic oscillator bath whose two-point
correlator is fixed to be $\left\langle \delta\hat{Q}(t)\delta\hat{Q}(0)\right\rangle $.
This approximation yields a long-time exponential decay at a rate
$\Gamma_{\varphi}=v^{2}/4\gamma$ for $T\gg\gamma$ (agreeing with
the results for classical telegraph noise, see above). At $T=0$,
one obtains a power law-decay $D(t)\sim t^{-\alpha}$ with an exponent
$\alpha=(4/\pi^{2})(v/\gamma)^{2}$, arising from the orthogonality
catastrophe \citep{anderson1967icf,HOPFIELD1969}. For larger coupling
strengths, $v/\gamma\gtrsim1$, the Gaussian approximation fails even
qualitatively, indicating the non-Gaussian nature of quantum telegraph
noise.

The important feature is the occurrence of visibility oscillations
beyond a critical coupling strength $v_{c}$. The visibility vanishes
at certain times and shows coherence revivals in-between. These features
continue to exist in the full quantum model. For $T\gg\gamma$, it
agrees with the classical result, 
where the threshold is $v_{c}^{cl}=\gamma$. In the quantum case (Fig.
\ref{fig:quantumcoherence}), we observe a transition to a non-monotonous
behaviour as a precursor to the visibility oscillations, in contrast
to the classical limit discussed above. Moreover, zeroes in the visibility
develop only at a larger coupling strength $v_{c}^{q}$, which depends
on temperature $T$. Another notable feature is the non-monotonous
evolution of peak heights for $v/\gamma \gtrsim 2.7$, unlike the classical
case. 

To illustrate these points, we have plotted the time-evolution of
$D(t)$ [excluding a trivial phase factor] as a function of the
coupling-strength $v$ for various temperatures {[}Fig. \ref{fig:densityplot}].
At high temperatures, visibility oscillations set in at $v_{c}^{q}/\gamma\approx1$,
whereas for $T\rightarrow0$ [Fig. \ref{fig:densityplot}(a)] the
first zero-crossing appears only at $v_{c}^{q}/\gamma\approx 2.7$.

\begin{figure}[t]
 \centering \includegraphics[width=0.99\columnwidth]{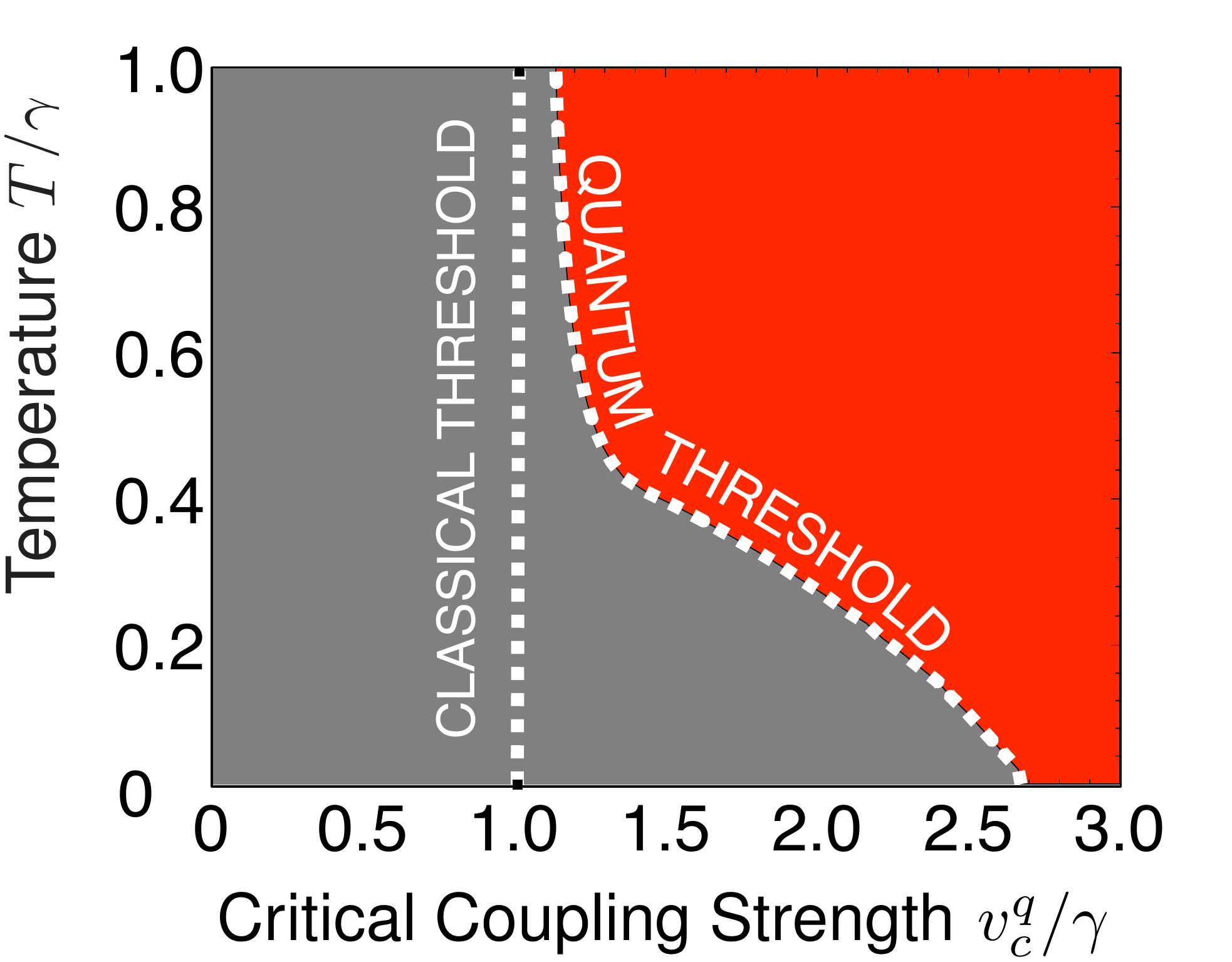}
\caption{Critical coupling strength $v_{c}^{q}(T)$ as a function of temperature
(with $\varepsilon_{0}=0,\gamma=1$). The strong-coupling regime is located
above the dashed line. At high temperatures, one has $v_{c}^{q}(T)\rightarrow1$,
according to the classical limit.}
\label{fig:phasediagram} 
\end{figure}

\textit{Temperature-dependence of strong-coupling threshold}.\textendash{}
As explained above, the visibility oscillations are a genuinely non-Gaussian
effect. We characterize the onset of the strong coupling regime by
the temperature-dependent critical coupling $v_{c}^{q}(T)$, beyond
which the zeroes in $D(t)$ appear. At a fixed temperature
$T$, the critical coupling-strength $v_{c}^{q}$ and the corresponding
zero in $D(t)$ at time $t^{*}$ are found numerically by a bisection
algorithm. The result is a {}``phase-diagram'' showing the critical
coupling $v_{c}^{q}$ as a function of $T$, [Fig.~\ref{fig:phasediagram}].
The curve $v_{c}^{q}(T)$ separates the $v-T$-plane into two regions.
At high temperatures $T$ the critical coupling $v_{c}^{q}$ converges
to its classical value, $v_{c}^{q}\rightarrow\gamma$ {[}a slight
offset in the plot is due to limited numerical accuracy]. For low $T$, it
increases and saturates at a finite value, as $D(t;v,T)$ is continuous
in the limit $T\rightarrow0$, and $D(t;v,T=0)$ still displays oscillations
beyond some threshold. This means the equilibrium quantum Nyquist
noise of the fluctuator is enough to observe visibility oscillations,
in contrast to the strong-coupling regime studied in  
\citep{2006_07_MZ_DephasingNonGaussianNoise_NederMarquardt},
where only the nonequilibrium shot noise of discrete electrons could
yield these effects.

\begin{figure}[t]
 \centering \includegraphics[width=0.99\columnwidth,keepaspectratio]{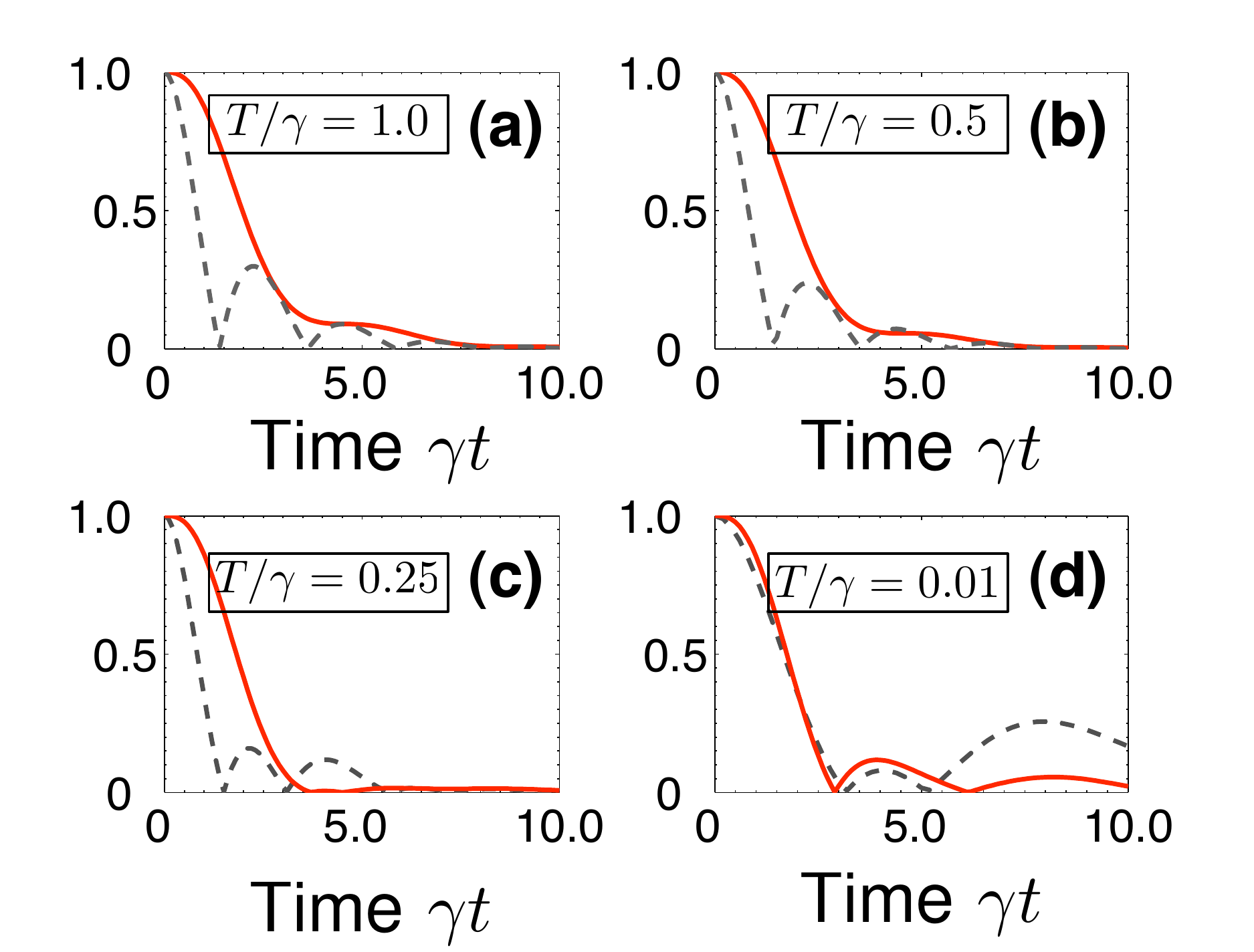} 
\caption{Time evolution of the spin-echo signal, $|D_{{\rm {Echo}}}(t)|$ (solid
line), after applying a $\pi$-pulse at $t' = t/2$, in comparison
with the visibility $|D(t)|$ for free evolution (dashed line), (a)-(d):
$T/\gamma=1.0$, $0.5$, $0.25$, $0.01$. ($v/\gamma=3.0$,
$\varepsilon_{0}=0$).}
\label{fig:echo} 
\end{figure}

\textit{Spin-Echo}.\textendash{} Finally, we investigate the time-evolution
of the density matrix of the charge qubit in a spin-echo experiment,
commonly employed to filter out low-frequency fluctuations, whose
effect is cancelled in such a procedure. Echo protocols were first
invented in NMR, but they are by now standard in qubit experiments,
particularly in the solid-state, where they are used to fight $1/f$
noise (\cite{Nakamura:2002xz}). At the initial time $t'=0$,
the qubit is prepared in a superposition of its two eigenstates, 
$\left|\psi(t_{0})\right\rangle =1/\sqrt{2}(\left|\uparrow\right\rangle +
\left|\downarrow\right\rangle )$. Then we let the qubit evolve according to 
Eq.~(\ref{eq:Hamiltonian}) up to a time $t'= t/2$, at which we perform a $\pi$-pulse
$e^{i\frac{\pi}{2}\hat{\sigma}_{x}}$ on the qubit, before evolving
up to time $t$. Defining $\hat{\mathcal{U}}_{\pm} = 
\exp[-i(\hat{\mathcal{H}}_{B}\pm v\hat{\mathcal{Q}}/2)t/2]$,
we find the qubit's final density matrix to be (in analogy to Eq.~(\ref{eq:densitymatrix})) $D_{{\rm {Echo}}}(t)=\left\langle \hat{\mathcal{U}}_{-}^{\dagger}\hat{\mathcal{U}}_{+}^{\dagger}\hat{\mathcal{U}}_{-}\hat{\mathcal{U}}_{+}\right\rangle$. 
As before, we can rewrite this as a determinant in the single-particle
Hilbert space, 
\begin{equation}
D_{{\rm {Echo}}}(t)=\det\left(1-\hat{n} + 
\hat{U}_{-}^{\dagger}\hat{U}_{+}^{\dagger}\hat{U}_{-}\hat{U}_{+}\hat{n}\right),
\end{equation}
where $\hat{U}_{\pm}$ is the single-particle evolution operator.
In Fig.~\ref{fig:echo} we compare the echo signal with the free
evolution. At low temperatures, the fluctuations are purely quantum
in origin, yielding a relatively lower weight for small frequencies
and thus a decrease in the effectiveness of the spin echo procedure.

\textit{Conclusion}.\textendash{} In conclusion, we have studied
the decoherence of a qubit subject to quantum telegraph noise. We
have calculated the time-evolution of the coherence and found a strong-coupling
regime with an oscillatory time-dependence of the coherence that cannot
be mimicked by any Gaussian noise source. We have characterized this
regime via the appearance of the first zero in the time-evolution
of the coherence and summarized the result in a ``phase-diagram''.
Moreover, we have presented the time-evolution of the echo-signal
in a spin-echo experiment and compared it to the coherence. Straightforward
extensions of the formulae presented here may be applied to discuss
the effects of more sophisticated pulse sequences  
 \citep{falci2004dst,gutmann2004bbr,santos2005dcq,uhrig2007kqb}
which are relevant for protecting quantum information storage.

\begin{acknowledgments}
We thank J.~Bergli and I.~Neder for useful discussions. We acknowledge the support through the DIP,  
the DFG via the SFB 631, the Nanosystems Initiative Munich (NIM),
the SFB/TR 12, and the Emmy-Noether program (F.~M.). 
\end{acknowledgments}

\end{document}